\documentclass[a4paper,10pt,oneside]{memoir}
\usepackage[magyar,english]{babel}
\usepackage{t1enc}
\usepackage[latin2]{inputenc}
\usepackage{graphics}
\usepackage{psfrag}
\usepackage[a4paper,left=2.5cm,right=2.5cm,top=2cm,bottom=2cm]{geometry}
\usepackage{amsmath}
\usepackage{amssymb}
\usepackage{amscd}
\usepackage{array}
\usepackage{enumerate}
\usepackage{tabularx}
\usepackage[mathscr]{eucal}
\usepackage{graphicx}
\usepackage{hhline}
\usepackage{wrapfig}
\usepackage{rotating}
\usepackage{verbatim}
\def\bibname{References}
\usepackage{ntheorem}
\usepackage{float}
\usepackage{subfig}
\usepackage{url}
\usepackage{authblk}

\usepackage[colorlinks=true, bookmarks=true]{hyperref}
\headstyles{bringhurst}
\newcommand{\R}{\mathbb{R}}

\newcommand{\er}{\left.\begin{aligned}}
\newcommand{\erv}{\end{aligned}\right\}}

\newtheorem{theorem}{Theorem}[section]

\newtheorem{notation}[theorem]{Notation}

\newtheorem{example}[theorem]{Example}

\newtheorem{remark}[theorem]{Remark}
\newtheorem{assumptions}[theorem]{Assumptions}

\graphicspath{{./figures/}}

\def\bibname{References}
\counterwithout{section}{chapter}

\title{Estimation of Claim Numbers in Automobile Insurance}
\author[1]{Miklós Arató}
\author[1]{László Martinek}
\affil[1]{Department of Probability Theory and Statistics, Eötvös Loránd University, Budapest}
\date{\today}

\begin{document}
\footnotetext[1]{The European Union and the European Social Fund have provided financial support to the project under the grant agreement no. TÁMOP 4.2.1/B-09/KMR-2010-0003.}

\maketitle
\numberwithin{figure}{section}
\numberwithin{equation}{section}

\pagenumbering{arabic}
\setcounter{page}{1}

\renewcommand{\contentsname}{Table of Contents}
\tableofcontents*
\setcounter{tocdepth}{1}

\begin{center}
\rule{4 cm}{0.01 cm}
\end{center}

\renewcommand{\abstractname}{Abstract}
\begin{abstract}
The use of bonus-malus systems in compulsory liability automobile insurance is a worldwide applied method for premium pricing. If certain assumptions hold, like the conditional Poisson distribution of the policyholders claim number, then an interesting task is to evaluate the so called claims frequency of the individuals. Here we introduce 3 techniques, two is based on the bonus-malus class, and the third based on claims history. The article is devoted to choose the method, which fits to the frequency parameters the best for certain input parameters. For measuring the goodness-of-fit we will use scores, similar to better known divergence measures. The detailed method is also suitable to compare bonus-malus systems in the sense that how much information they contain about drivers.

\vspace{.3 cm}
\noindent
Keywords: bonus-malus, claims frequency, Bayesian, scores
\end{abstract}

\section[Introduction]{Introduction}
\label{sec:intro}
\subsection[Motivation]
{Motivation}
\label{subsec:motiv}
In compulsory liability automobile insurance a worldwide applied method is to calculate the premium of the drivers based on a so-called bonus-malus system. This might be vary for different countries, but the idea is similar: policy-holders with bad history should pay more than others without accidents in the past years. Schematically and mathematically described, there is a graph with certain vertices (they called classes), among others an initial vertex, where every new driver begins. After a year without causing any accident he or she jumps up to another class, which has a cheaper premium. Otherwise, in case of causing an accident, the insured person goes downward, to a new class with higher premium, except he or she was already in the worst one.

Also many other factors are taken into consideration when calculating ones premium, like the engine type, purpose of use, habitat, age of the person etc. Here we do not deal with these components, only concentrate on the bonus-malus system. (There are many famous books about bonus-malus systems where we can read about these factors, e.g. \cite{Lemaire1996}.) Namely, our aim is to estimate the expected $\lambda$ number of accidents for certain drivers. This is usually called the claims frequency of the policyholder.  First we review our necessary assumptions, among others about the distribution of claim numbers of a policyholder in one year, and the Markovian property of a random walk in such a system. Section \ref{sec:bayes} will discuss the basic problem illustrated with Belgian, Brazilian and Hungarian example. Since also a Bayesian approach will be used for estimation of $\lambda$, a general a priori assumption is also needed. Based on this and a gappy information about the driver, the a posteriori expected value of $\lambda$ will be calculated.

Giving the possible closest estimation for $\lambda$ to the actual one is a crucial task in the insurers life, since the expected value of claims, and which is the most important, the claims payments are forecasted using $\lambda$. Let us note that for evaluating the size of payments on the part of the insurer, the size of the certain property damages has to be approximated, not even the number of them. Here we will not care for it in this article.

\subsection[Bonus-malus systems]
{Bonus-malus systems}
\label{subsec:bms}

We suppose the following assumptions, however, it is surely a simplification of the reality:
\begin{assumptions}
\begin{enumerate}[$\bullet$]
\item $\lambda$ is a constant value in time \footnote{The case of the variable $\lambda$ in time can be handled using double stochastic processes (Cox processes for instance), i.e. $\lambda(t)$ would also be a stochastic process.}
\item the random walk on the graph of classes is a homogeneous Markov chain, i.e. the next step depends only on the last state, and independent of time
\item the distribution of a policyholders claim number for a year is Poisson($\lambda$) distributed
\end{enumerate}
\end{assumptions}

\begin{notation}
For a bonus-malus system consisting of $n$ classes, let $C_1$ denote the worst one with the highest premium, $C_2$ the second worst etc., and finally $C_n$ the best premium class which can be achieved by a driver. On the other hand, let $B_t$ be the class of the policy-holder after $t$ steps (years).
\end{notation}

Using these notations, the Markovian property can be written as $P(B_t = C_i | B_{t-1},\ldots,B_1) = P(B_t=C_i | B_{t-1})$. As the $B_t$ process is supposed to be homogeneous, it is correct to simply write $p_{ij}$ instead of $P(B_{n+1} = j | B_n = i)$. These values specify an $n \times n$ stochastic matrix with non-negative elements, namely the transition probability matrix of the random walk on states. Let us denote it by $M(\lambda)$. Now to be more specific, we outline the example of the three different systems, the Belgian, Brazilian and Hungarian.

\begin{example}[Hungarian system]
In the Hungarian bonus-malus system there are 15 premium classes, namely an initial ($A_0$), 4 malus ($M_4, \ldots, M_1$) and 10 bonus ($B_1, \ldots, B_{10}$) classes\footnote{We have to be careful here not to confuse the class index of $B_i$ with the time index!}. In our above mentioned terminology, we can think of it as $C_1 = M_4, \ldots, C_4 = M_1, C_5 = A_0, C_6 = B_1, \ldots, C_{15} = B_{10}$. After every claim-free year the policyholder jumps one step up, unless he or she was in $B_{10}$, when there is no better class to go. The consequence of every reported damage is 2 classes relapse, and at least 4 damages pulls the driver back to the worst $M_4$ state. Thus the transition probability matrix takes the form of Equation \ref{eq:transmtrx_hu}, see Appendix.
\end{example}

\begin{example}[Brazilian system]
7 premium classes: $A_0, B_1, B_2, \ldots, B_6$. Sometimes written as 7, 6, 5, $\ldots$, 1 classes, e.g. in Jean Lemaire's article \cite{Lemaire1998}. Transition rules can be found in the cited article.
\end{example}

\begin{example}[Belgian system]
Here we look at the new Belgian system introduced in 1992, and within this we focus on business-users. The only difference between them and pleasure-users is the initial class. Transition rules can also be found in article \cite{Lemaire1998}. There are 23 premium classes: $M_8$,  $M_7$,  $\ldots$,  $M_1$,  $A_0$,  $B_1$,  $B_2$, $\ldots, B_{14}$ (sometimes written as $23, 22, 21, \ldots, 2, 1$ classes).
\end{example}
: The introduction  is still missing! \par}

\section[The Bayesian approach]{The Bayesian approach}
\label{sec:bayes}

Our realistic problem is the following. When an insured person changes insurance institution, the new company not necessarily gets his or her claim history, only the class where his or her life has to keep going. The new insurer also knows the number of years the policyholder has spent in the liability insurance system. Nevertheless, based on this two information we would like to provide the best possible estimation for the certain persons $\lambda$. First of all we need some new notations.

\begin{notation}
$B_t = c$ denotes the event that the investigated policyholder has spent $t$ years in the system (more precisely, from the initial class he or she has taken $t$ steps), and arrived in class $c$.
\end{notation}

\begin{notation}
$\pi_0$ is the initial discrete distribution on the graph, which is a row vector of the form \\ $(0, \ldots, 0,\underbrace{1}_{i^{th}},0, \ldots, 0)$. (Every driver begins in the initial $C_i$ state.)
\end{notation}

As a Bayesian approach, suppose that claims frequency is also a random variable, and denote it by $\Lambda$. In practice, the gamma mixing distribution seems to be a right choice for that, therefore only this case will be discussed now. For other cases, the calculations can be similarly, though not similarly nicely done.

\begin{notation}
$\Gamma(\alpha,\beta)$ is the gamma distribution with $\alpha$ shape and $\beta$ scale parameters. (Defining the scale parameter precisely, the density function is $f(x) = \frac{x^{\alpha-1} \cdot \beta^\alpha \cdot e^{-\beta x}}{\Gamma (\alpha)}$.)
\end{notation}

For some fixed $\Lambda = \lambda$ the driver causes $X \sim $Poisson($\lambda$) property damages, i.e. $P(X=k | \Lambda = \lambda) = \frac{\lambda^k}{k!} e^{-\lambda}$. It is well known that the unconditional distribution will be negative binomial with parameters $\left( \alpha , \frac{\beta}{1+ \beta} \right)$ in our notations. In view of this the a priori parameters can be estimated by a standard momentum or maximum likelihood method.

\begin{remark}
It is certainly necessary to make hypothesis testing after all, because our assumptions about the mixing distribution might be wrong. In this negative-binomial-rejecting case we have two important alternatives. They are to be found among others in \cite{Fishman1996}, for instance. On the one hand we can try inverse Gaussian, and on the other hand lognormal distribution for mixing.
\end{remark}

\subsection[Estimation of distribution parameters]
{Estimation of distribution parameters}
\label{sec:estimpar}

In this subsection we give an estimation method to compute the approximate values of $\alpha$ and $\beta$ parameters. Let us suppose that the insurance company has claim statistics from the past few years containing $n$ policyholders. The $i$th insured person caused $X_i$ accidents by his or her fault over a time period of $t_i$ years, where $t_i$ is not necessarily an integer. According to our assumption, the distribution of $X_i$ is Poisson($t_i \cdot \Theta$), where $t_i$ is a personal time factor and $\Theta$ is a Gamma($\alpha, \beta$)-distributed random variable. The unconditional distribution of $X_i$ as mentioned above is Negative Binomial$\left( \alpha, \frac{\beta}{t_i + \beta} \right)$, thus its first two moments are
\begin{equation}
E X_i = t_i \frac{\alpha}{\beta}
\end{equation}
\begin{equation}
E X_i^2 = t_i^2 \frac{\alpha}{\beta^2} + t_i^2 \frac{\alpha^2}{\beta^2} + t_i \frac{\alpha}{\beta}.
\end{equation}
Now we construct a method of moments estimation, which is not obvious how to build. For example, on the one hand, we can say that $\sum \limits_{i=1}^n EX_i = \frac{\alpha}{\beta} \sum \limits_{i=1}^n t_i$, thus $\frac{\hat{\alpha}}{\hat{\beta}} = \frac{\sum \limits_{i=1}^n x_i}{\sum \limits_{i=1}^n t_i}$. On the other hand, $\sum \limits_{i=1}^n \frac{EX_i}{t_i} = n \frac{\alpha}{\beta}$, thus $\frac{\hat{\alpha}}{\hat{\beta}} = \frac{\sum \limits_{i=1}^n \frac{x_i}{t_i}}{n}$. Generally they do not give the same results, except $t_1 = t_2 = \ldots = t_n$. Based on our simulations, we used the most accurate method, where the approximation of parameters are the result of the following equations.
\begin{equation}
\frac{\hat{\alpha}}{\hat{\beta}} = \frac{\sum \limits_{i=1}^n X_i}{\sum \limits_{i=1}^n t_i}
\end{equation}
\begin{equation}
\frac{1}{\hat{\beta}} = \frac{\sum \limits_{i=1}^n t_i}{\sum \limits_{i=1}^n t_i^2} \left( \frac{\sum \limits_{i=1}^n X_i^2}{\sum \limits_{i=1}^n X_i} - 1 \right) - \frac{\sum \limits_{i=1}^n X_i}{\sum \limits_{i=1}^n t_i}
\end{equation}
Finally we have to notice the maximum likelihood method as another obvious solution for this parameter estimation. Unfortunately, the likelihood function generally has no maximum, hence this method has been rejected.

\subsection[Conditional probabilities]
{Conditional probabilities}
\label{sec:condprob}

\begin{figure}
  \centering
  \subfloat[Brazilian, $\alpha=1.2, \beta=19$]{\label{fig:bra1}\includegraphics[width=0.4\textwidth]{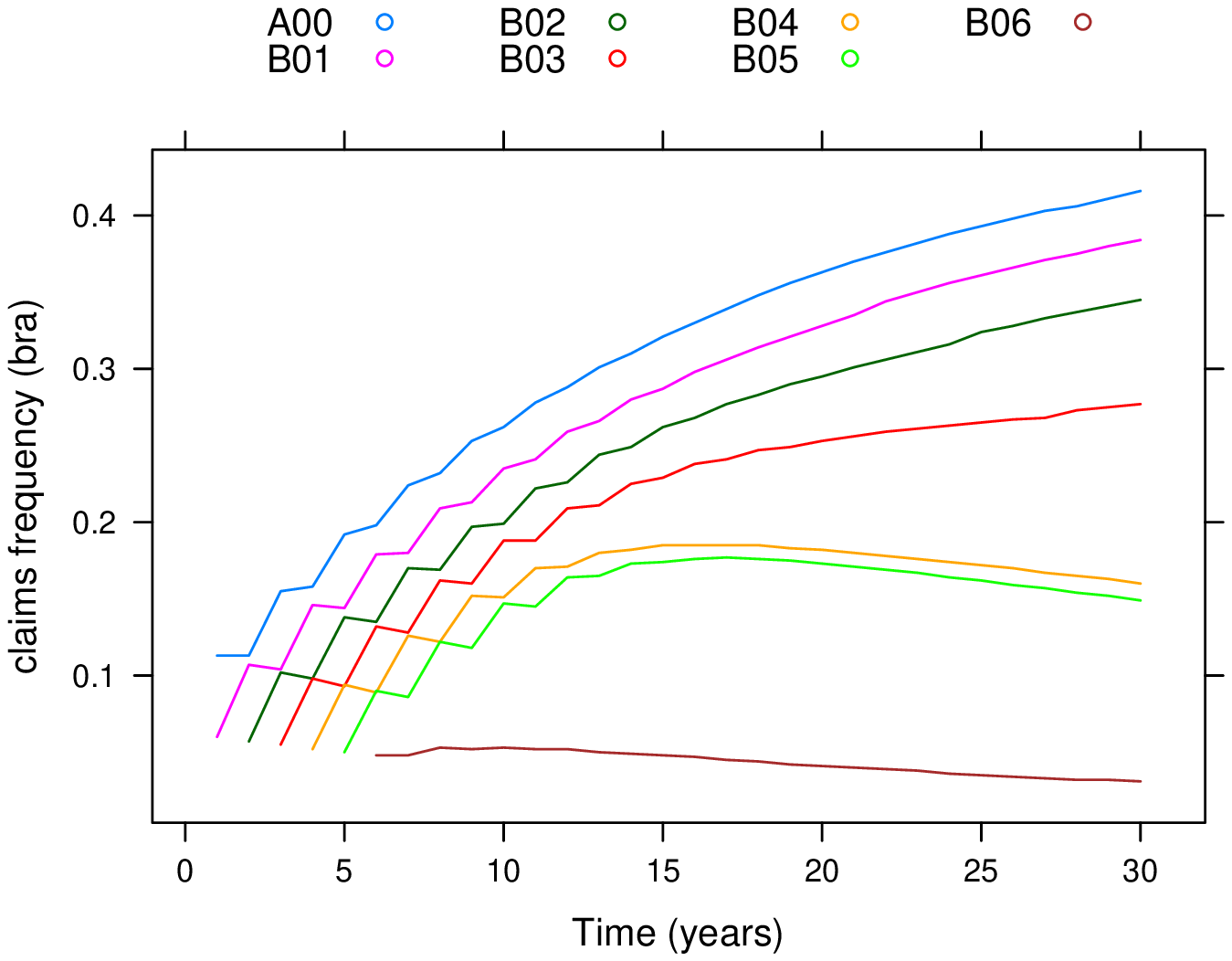}}
  \subfloat[Brazilian, $\alpha=1.8, \beta=12$]{\label{fig:bra2}\includegraphics[width=0.4\textwidth]{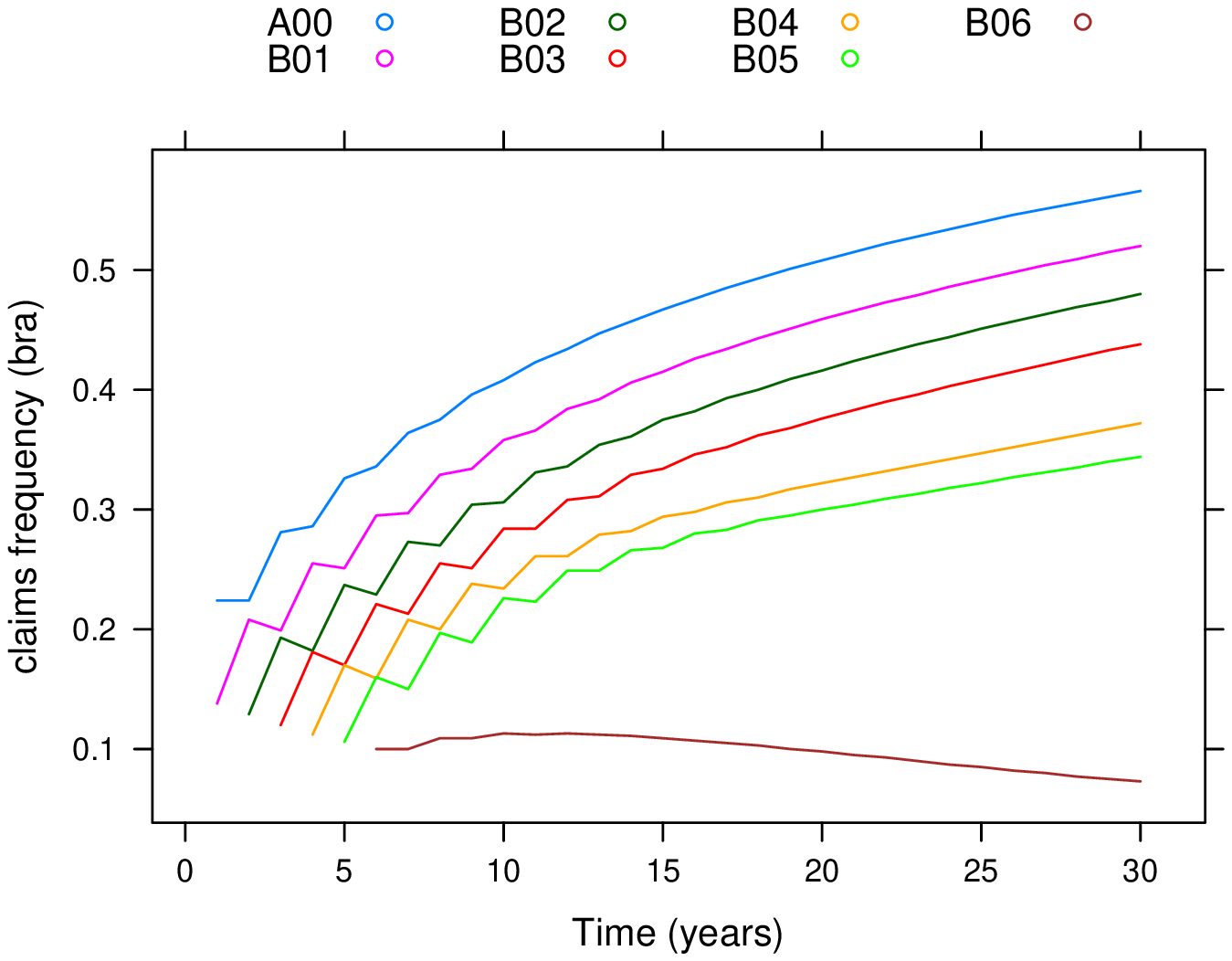}} \\
  \subfloat[Hungarian, $\alpha=1.2, \beta=19$]{\label{fig:hun1}\includegraphics[width=0.4\textwidth]{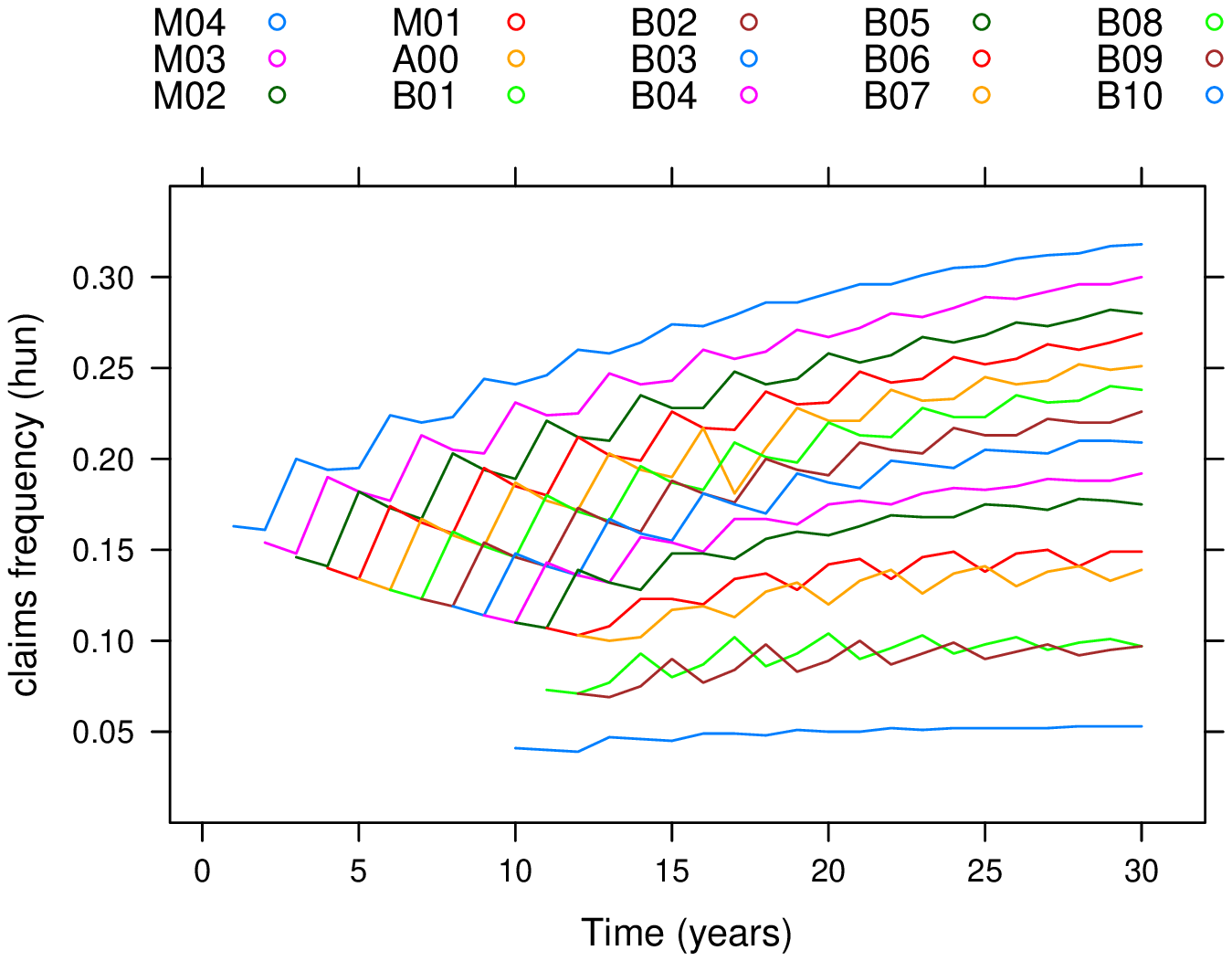}}
  \subfloat[Hungarian, $\alpha=1.8, \beta=12$]{\label{fig:hun2}\includegraphics[width=0.4\textwidth]{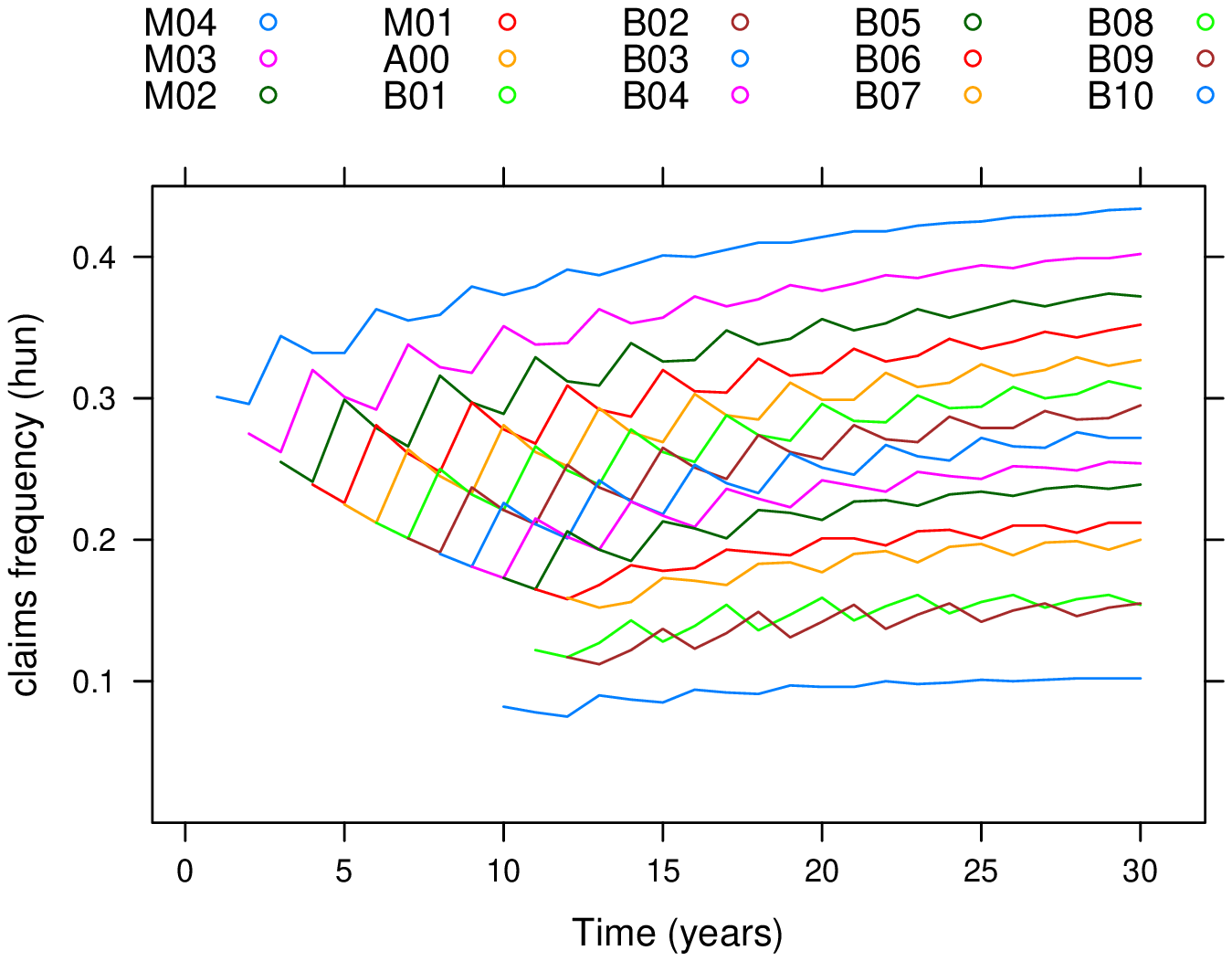}} \\
  \subfloat[Belgian, $\alpha=1.2, \beta=19$]{\label{fig:bel1}\includegraphics[width=0.4\textwidth]{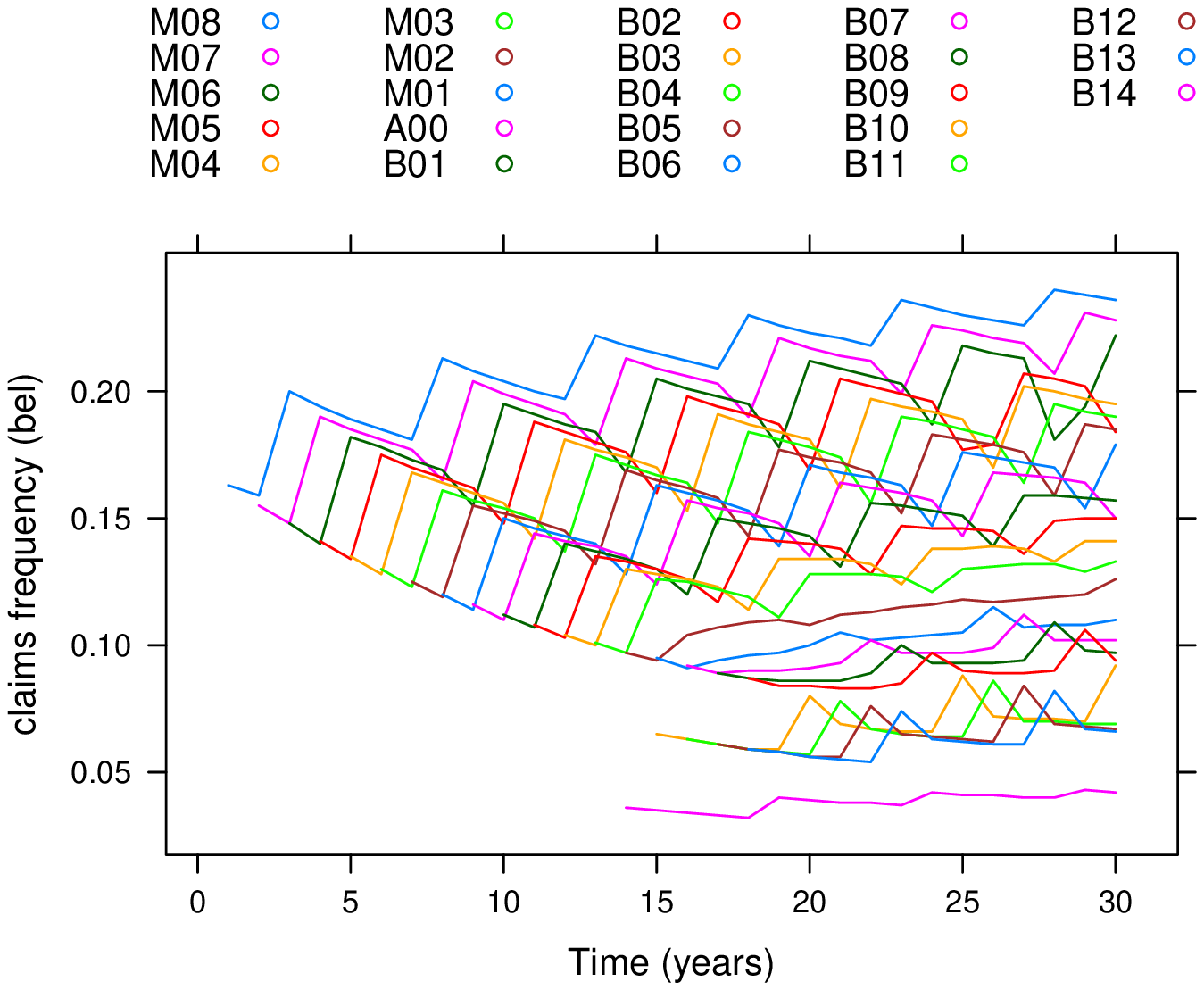}}
  \subfloat[Belgian, $\alpha=1.8, \beta=12$]{\label{fig:bel2}\includegraphics[width=0.4\textwidth]{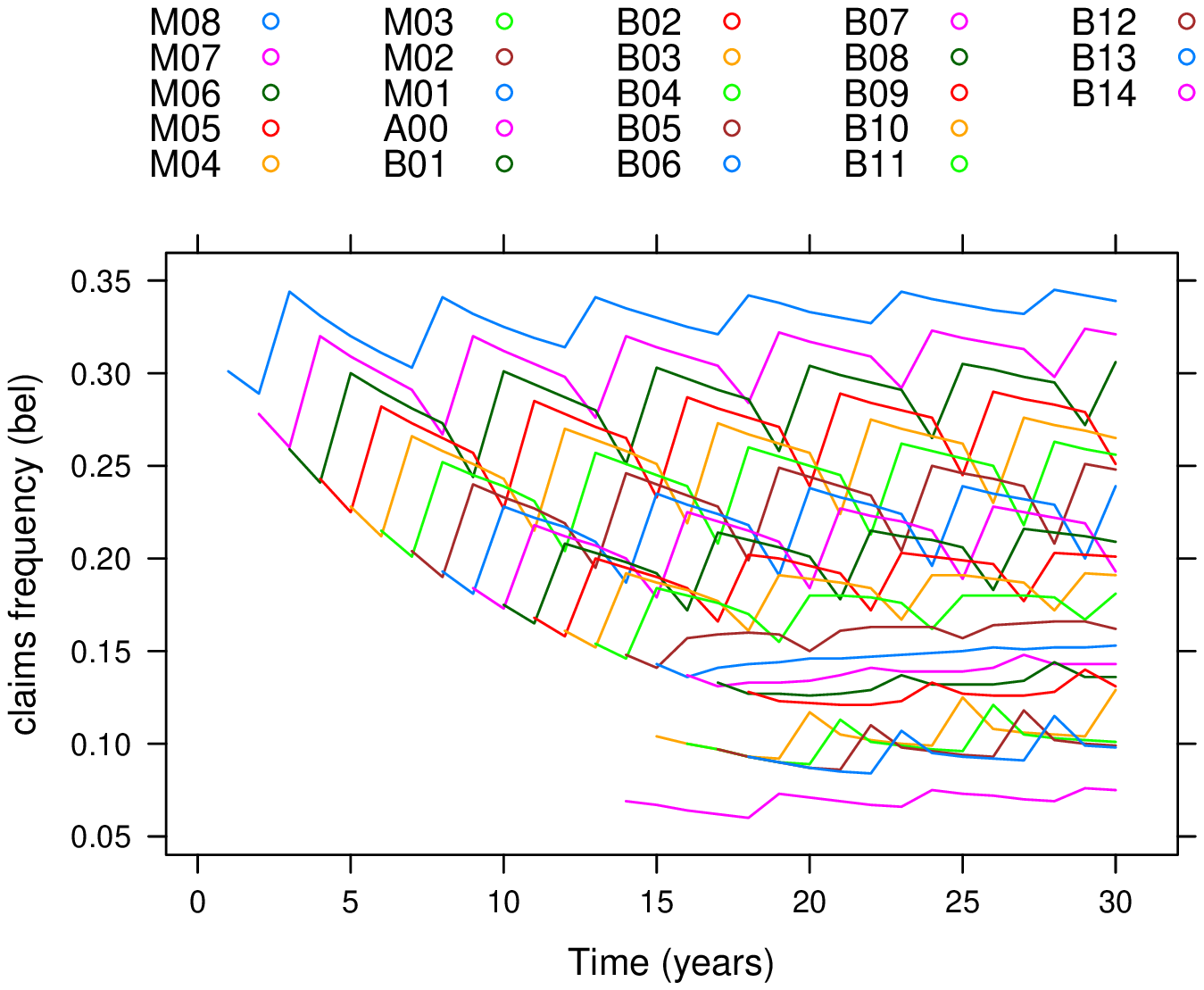}} \\

  \caption{Estimated $\lambda$ parameters for different countries and $\alpha, \beta$ parameters on a time horizon of 30 years. (Note that some points in the colored version are not shown in these figures. For every class, charts start at points, from where the probability of being there is positive.)}
  \label{fig:estim1}
\end{figure}

Now we have the a priori $\alpha, \beta$ parameters, and the $B_t = c$ information with the initial class $=B_0$. According to Bayes theorem, the conditional density of $\Lambda$ is $$f_{\Lambda | B_t=c} (\lambda|c) = \frac{P(B_t=c | \Lambda = \lambda) \cdot f_\Lambda (\lambda)}{\int \limits_0^\infty P(B_t=c | \Lambda = \lambda) \cdot f_\Lambda (\lambda) d\lambda},$$ and the estimation for $\lambda$ is the a posteriori expected value $$\hat{\lambda} \stackrel{\text{not.}}{=}  \int \limits_0^\infty \lambda \cdot f_{\Lambda | B_t=c} (\lambda|c) d\lambda.$$ Unfortunately, $P(B_t=c | \Lambda = \lambda)$ presents some difficulty, because it only can be calculated pointwise as a function of $\lambda$, powering the $M(\lambda)$ transition matrix. If $I$ denotes the index of the initial class in the graph, this probability is exactly the $M(\lambda)^t_{(I,|c|)}$ element of the matrix, where $|c|$ is the index of class $c$. Using numerical integration it can be solved relatively fast. If we have a glance at figure \ref{fig:estim1}, we might see the claim frequency estimations for different countries and different $\alpha$ and $\beta$ parameters. For example, the first figure \ref{fig:bra1} was made based on the Brazilian bonus-malus system with parameters $\alpha = 1.2$ and $\beta = 19$, which indicates a relatively high risk portfolio. There is a line in the chart for each bonus class, which values show the estimated $\lambda$ frequencies in function of time. We also see that these functions stationarize in more decades, thus the information about in the system elapsed time of the individual is important.

We will refer to this method as \texttt{method.1} later.

: The Bayesian approach section  is still missing! \par}

\section[Other methods for frequency estimation]{Other methods for frequency estimation}
\label{sec:meth23}

In this section we shortly introduce 2 other (known) possible methods to evaluate policyholders' claim frequencies. Our assumptions for the distribution of claim numbers certainly still hold.

\subsection[Average claim numbers of classes]
{Average claim numbers of classes}
\label{subsec:meth2}

The following method is probably the simpliest, and we will refer to this as \texttt{method.2}. Let us suppose that statistics from the last year claim numbers are available. Take the average caused claims for every bonus classes, i.e., if last year our portfolio contained 5 policyholders each in class $C$ with claims $0,1,0,0,2$, then the estimation for insured people's frequencies present year in class $C$ is $\hat{\lambda}_C = 0.6$. Although it seems to be too simplified, in some cases it gives the best results in certain circumstances.

\subsection[Claim history of individuals]
{Claim history of individuals}
\label{subsec:meth3}

At last the third method is classical, and will be referred as \texttt{method.3}. Here we use the insured persons claim history, i.e., suppose that he or she was insured by our company for $t$ years, and the distribution of claim numbers for each year is Poisson($\lambda$). Let $X$ denote the number of aggregate claims caused by this person in his $t$-year-long presence in the system, more precisely, in our field of vision. Then the conditional distribution of $X$ is also Poisson with parameter $\lambda t$. Let us remark that $t$ does not have to be an integer (people often change insurer in the middle of the year in most countries). Based on this, our estimation on a certain $\lambda$ is the conditional expected value of the Gamma distributed $\Lambda$ on condition $X=x$, i.e., as well known, $\hat{\lambda} = E(\Lambda | X=x) = \frac{x+\alpha}{t+\beta}$. 

This is also a Bayesian approach as in the case of \texttt{method.1}, but the condition is different. Besides notice that first the $\alpha$ and $\beta$ parameters have to be estimated exactly the way we have seen it before in Section \ref{sec:estimpar}. : The Other methods for frequency estimation section  is still missing! \par}

\section[Comparison using scores]{Comparison using scores}
\label{sec:score}

The aim of this section is to make a decision which method gives the possible most accurate estimation for claims frequencies. In this context, we will use the theory of so called scores. For much more detailed information see article \cite{Gneiting2007}, as here we will discuss only the most important properties useful for our problem.

Scores are made for measuring the accuracy of probabilistic forecasts, i.e., measuring the goodness-of-fit of our evaluations. Let $\Omega$ be a sample space, $\mathcal{A}$ a $\sigma$-algebra of subsets of $\Omega$ and $\mathcal{P}$ is a family of probability measures on $(\Omega, \mathcal{A})$. Let a scoring rule be a function $S : \mathcal{P} \times \Omega \longrightarrow \overline{\R} = [-\infty, \infty]$. We work with the expected score $S(P,Q) = \int S(P,\omega) dQ(\omega)$, where measure $P$ is our estimation and $Q$ is the real one. Obviously one of the most important properties of this function is the inequality $S(Q,Q) \geq S(P,Q)$ for all $P,Q \in \mathcal{P}$. In this case $S$ is proper relative to $\mathcal{P}$. On the other hand, we call $S$ regular relative to class $\mathcal{P}$, if it is real valued, except possibly it can be $-\infty$ in case of $P \neq Q$. If these two properties hold, then the associated divergence function is $d(P,Q) = S(Q,Q) - S(P,Q)$.

Here we mention two main scoring rules, which will be used in sections below for comparing our evaluation methods. Remember that for an individual, the conditional distribution of the number of accidents is Poisson, so in our notations let $p_i$ ($i=0,1,2,\ldots$) be $P(X = i | \Lambda = \hat{\lambda})$, i.e., the probabilities of certain claim numbers using the estimated $\hat{\lambda}$ as condition. Similarly, $q_i$ ($i=0,1,2,\ldots$) is the same probability, but for the real $\lambda$ frequency.

\subsection[Brier score]
{Brier Score}
\label{subsec:brier}

Sometimes called quadratic score, as the associated Bregman divergence is $d(p,q) = \sum \limits_i (p_i - q_i)^2$. The score is defined as
\begin{equation}
S(P,Q) = 2 \left( \sum \limits_i p_i q_i \right) - \left( \sum \limits_i p_i^2 \right) - 1.
\end{equation}
Here we use deliberately the unknown $q_i$ probabilities. Although, in practice they are not know, in our simulations we are able to make decisions using them. If we want to calculate the score of our estimation subsequently, we change $q_i$s to a Dirac-delta depending on the number of claims caused. In other words, if the examined policyholder had $i$ claims last year, then the corresponding score is $$S(P,i) = 2 \left( \sum \limits_j p_j \delta_{ij} \right) - \left( \sum \limits_j p_j^2 \right) - 1 = 2p_i - \sum \limits_j p_j^2 - 1.$$

\subsection[Logarithmic score]
{Logarithmic Score}
\label{subsec:logarithmic}

The logarithmic score is defined as
\begin{equation}
S(P,Q) = \sum \limits_i q_i \log p_i.
\end{equation}
(In case of $i$ caused accidents $S(P,i) = \log p_i$.) We mention that the associated Bregman divergence is the Kullback-Leibler divergence $d(p,q) = \sum \limits_i q_i \log \frac{q_i}{p_i}$. : The Comparison using scores section  is still missing! \par}

\section[Simulation and results]{Simulation and results}
\label{sec:simul}

Using R program we simulated portfolios for testing our methods the following way. This can be used for frequency evaluation in practice, if we have the appropriate inputs. First we need a portfolio containing $N$ insured people, which is used to estimate the $\alpha$ and $\beta$ parameters of the negative binomial distribution. We think of it as the policyholders' histories available in the insurers database. Taking advantage of the whole claim and bonus-malus history, we have done it exactly the way as described in Section \ref{sec:estimpar}. In the next step, we generated the history of an $M$ policyholder-containing portfolio, assuming that the distribution parameters are unchanged compare to the first portfolio. This might result some bias, but this is the best we can do based on our available data.

After that, we estimated the claim frequency parameters of policyholders separately based on the three estimation methods described above, and compared them to the real $\lambda$ parameters using scores. Our aim is to make a decision among the methods, and decide, which would give the best fit results in certain cases. In other words, for certain input parameters, which method gives good-fit estimations in expected value, where the higher score value means the better goodness-of-fit. Before discussing results, we better stop for a few remarks.

As we are interested in the expected values of scores for certain inputs, we will apply a Monte-Carlo-type technique. This means that we generate the above mentioned two portfolios $n$ times independently, but in each first ones preserving the $\alpha$ and $\beta$ distribution inputs. After that, based on the approximated $\hat{\alpha}$ and $\hat{\beta}$, we estimate the $\lambda$ parameters of the second portfolios. Each method gives one score number for each simulation, which is the average of scores calculated for individuals. (Of course aggregate scores would be equally appropriate, since it differs only in an $M$ multiplier from the average.) At last we take the mean of mean scores, and the higher score resulting method is the better.

\begin{remark}
For reference we will write and plot the score results also for comparing the real frequencies to the real frequencies, since these scores are not equal to 0, as in the divergence case. In function of year steps, these scores should be constant, contrary to the charts below, where small differences can be observed. The reason is that in the Monte-Carlo simulations we generated also the $\lambda$ parameters over and over.
\end{remark}

\begin{remark}
In our simulations input parameters are
\begin{enumerate}
\item $N$ the number of policyholders in the first portfolio, which is used to estimate the $\alpha$ and $\beta$ parameters.
\item $M$ the number of policyholders in the second portfolio. This contains individuals, whose $\lambda$ parameters have to be evaluated. In practice, there might be an overlap between these two files.
\item Real $\alpha$ and $\beta$ distribution parameters.
\item Number of year steps. This means the time elapsed in years, since the certain individual is insured by our company. Note that it implies the knowledge of claims history and bonus classification, thus it is a very important feature, as it affects the goodness-of-fit of our estimation methods.
\item Number of years elapsed before entering our company. This affects \texttt{method.1} and \texttt{method.2}, because the Markov chain on the bonus classes converges slowly to the stationary distribution.
\item Transition rules of the examined country.
\item Number of simulated portfolios. As we approximate the scores via Monte-Carlo-type technique, it has to be large enough.
\end{enumerate}
\end{remark}

\begin{figure}
  \centering
  \subfloat[Brazilian, Brier scores]{\includegraphics[width=0.4\textwidth]{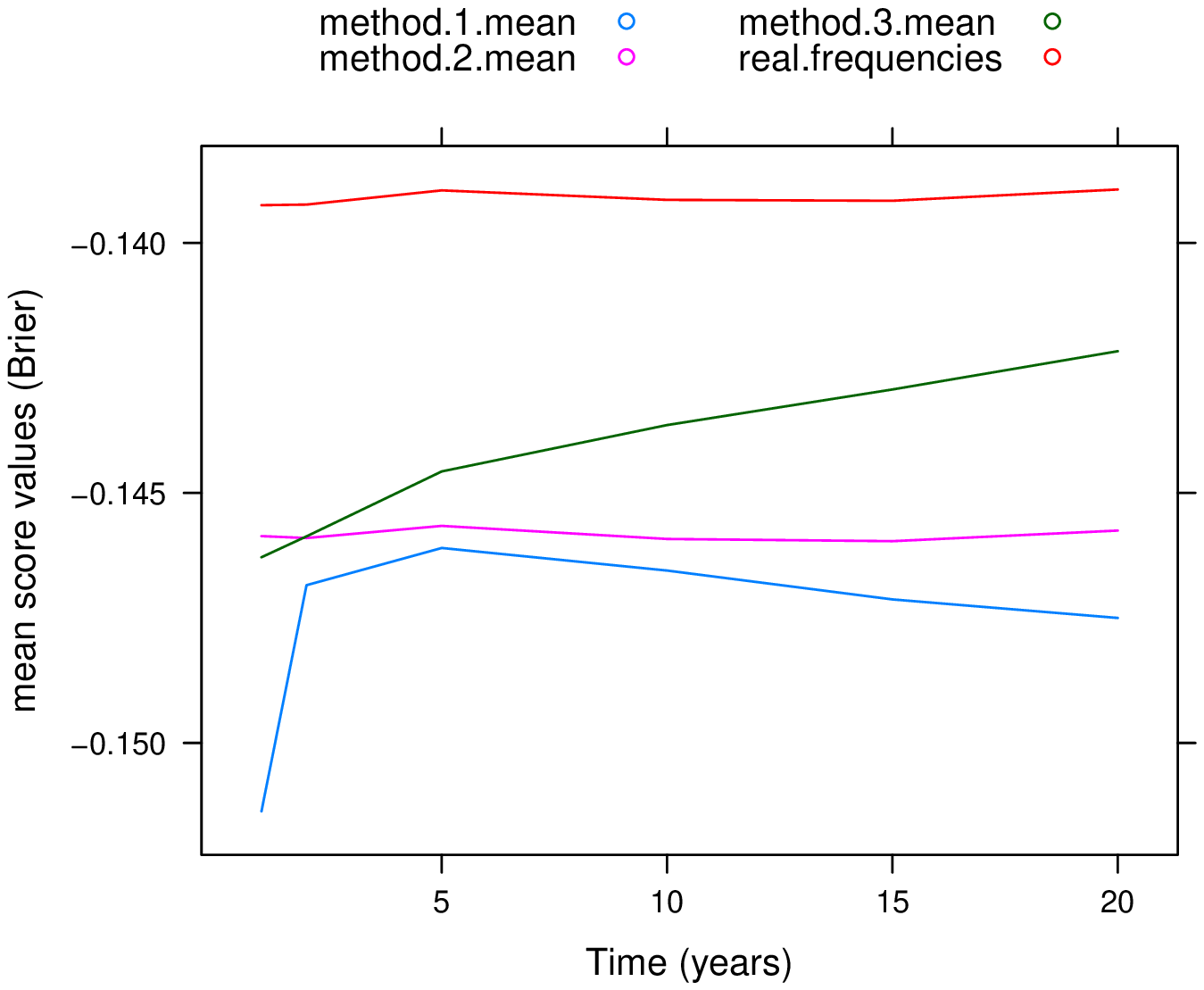}}
  \subfloat[Brazilian, Log scores]{\includegraphics[width=0.4\textwidth]{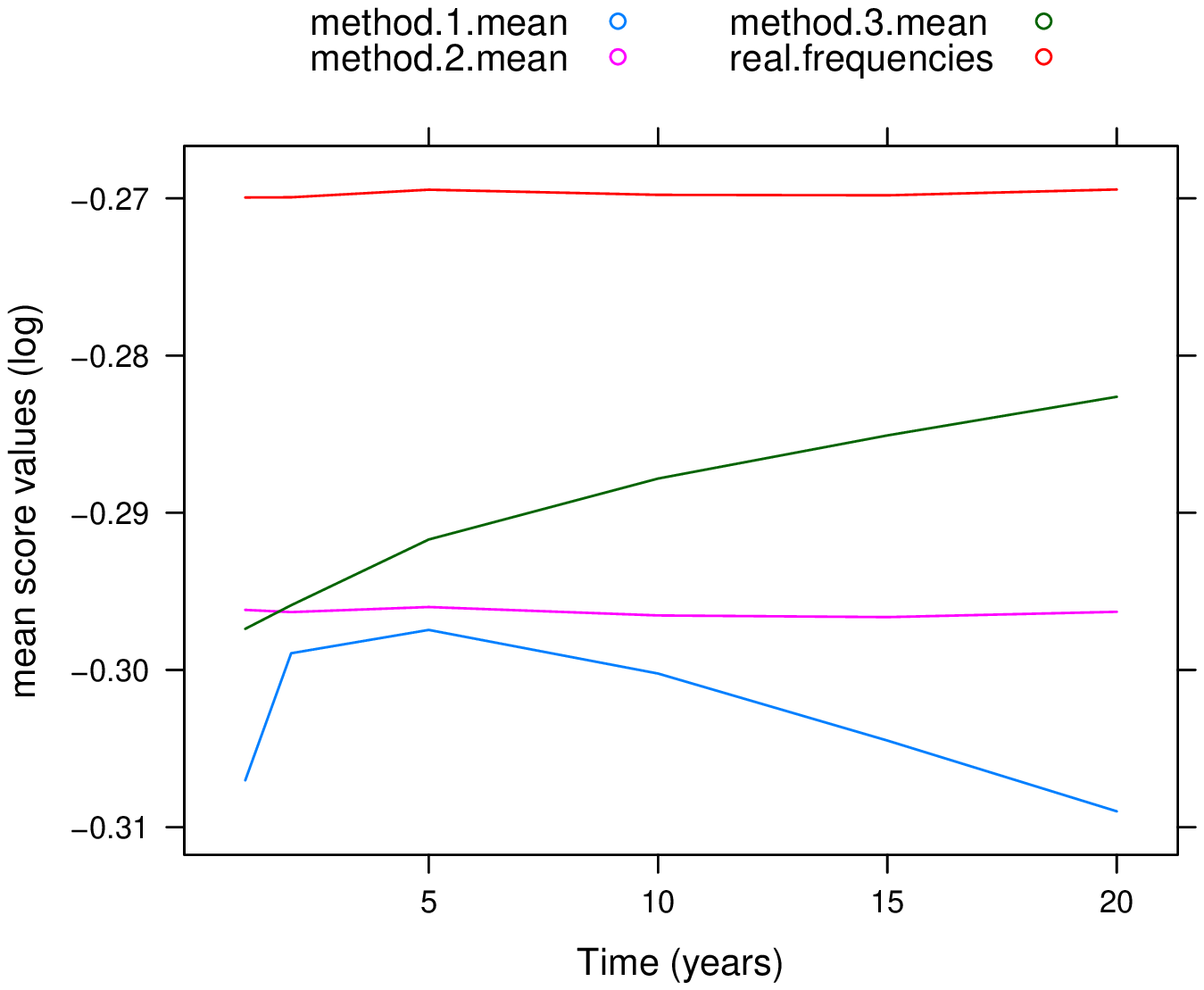}} \\
  \subfloat[Hungarian, Brier scores]{\includegraphics[width=0.4\textwidth]{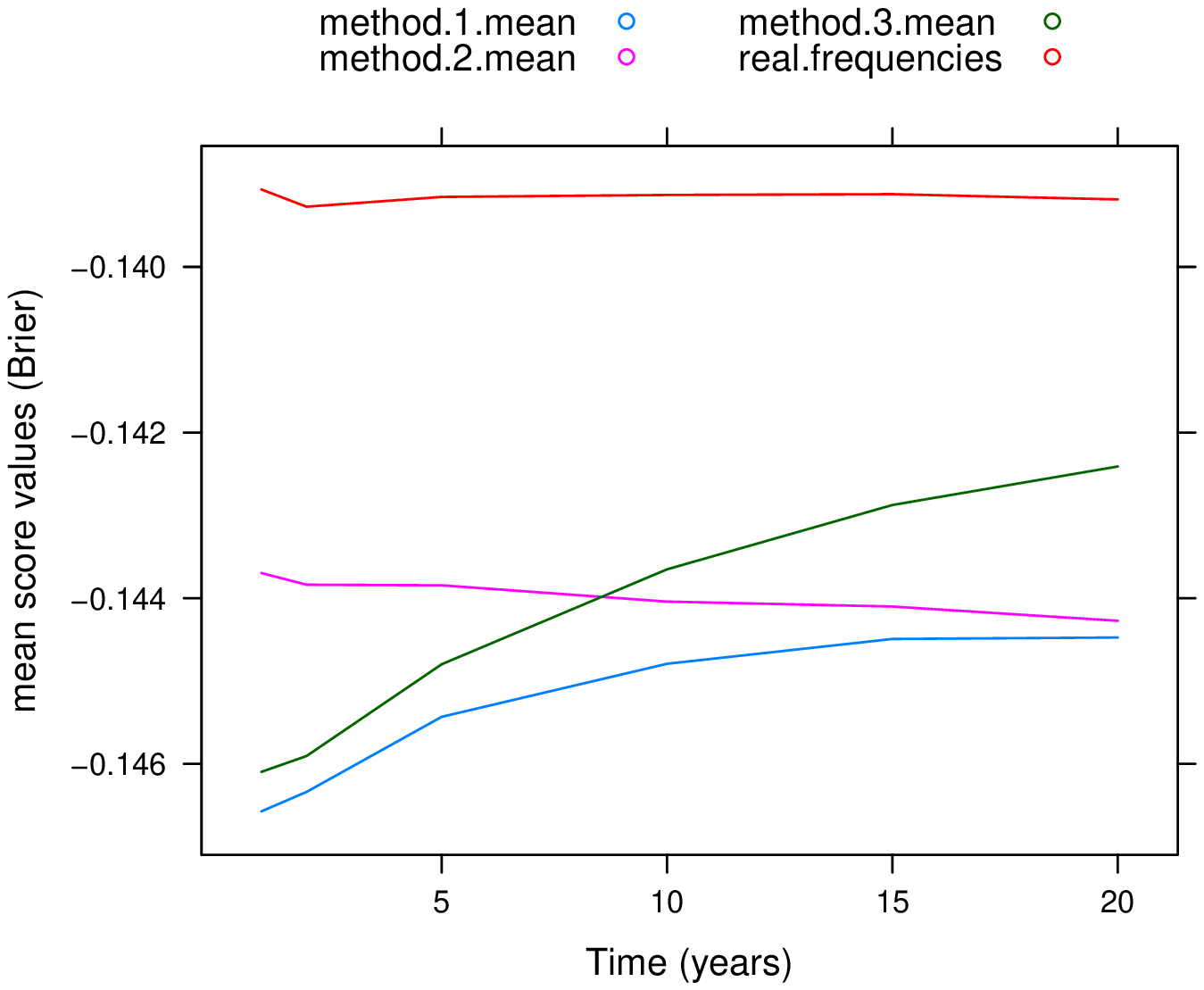}}
  \subfloat[Hungarian, Log scores]{\includegraphics[width=0.4\textwidth]{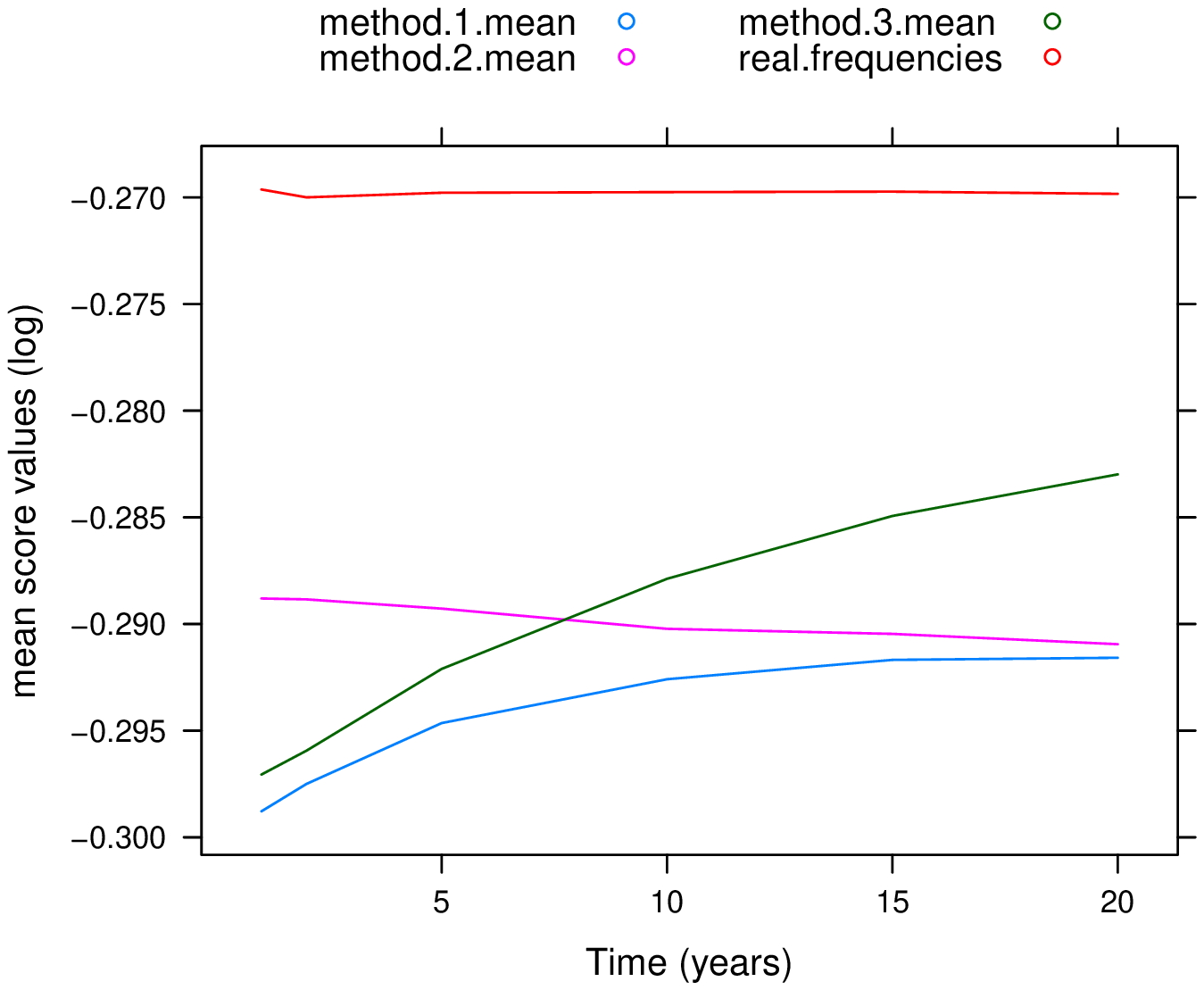}} \\
  \subfloat[Belgian, Brier scores]{\includegraphics[width=0.4\textwidth]{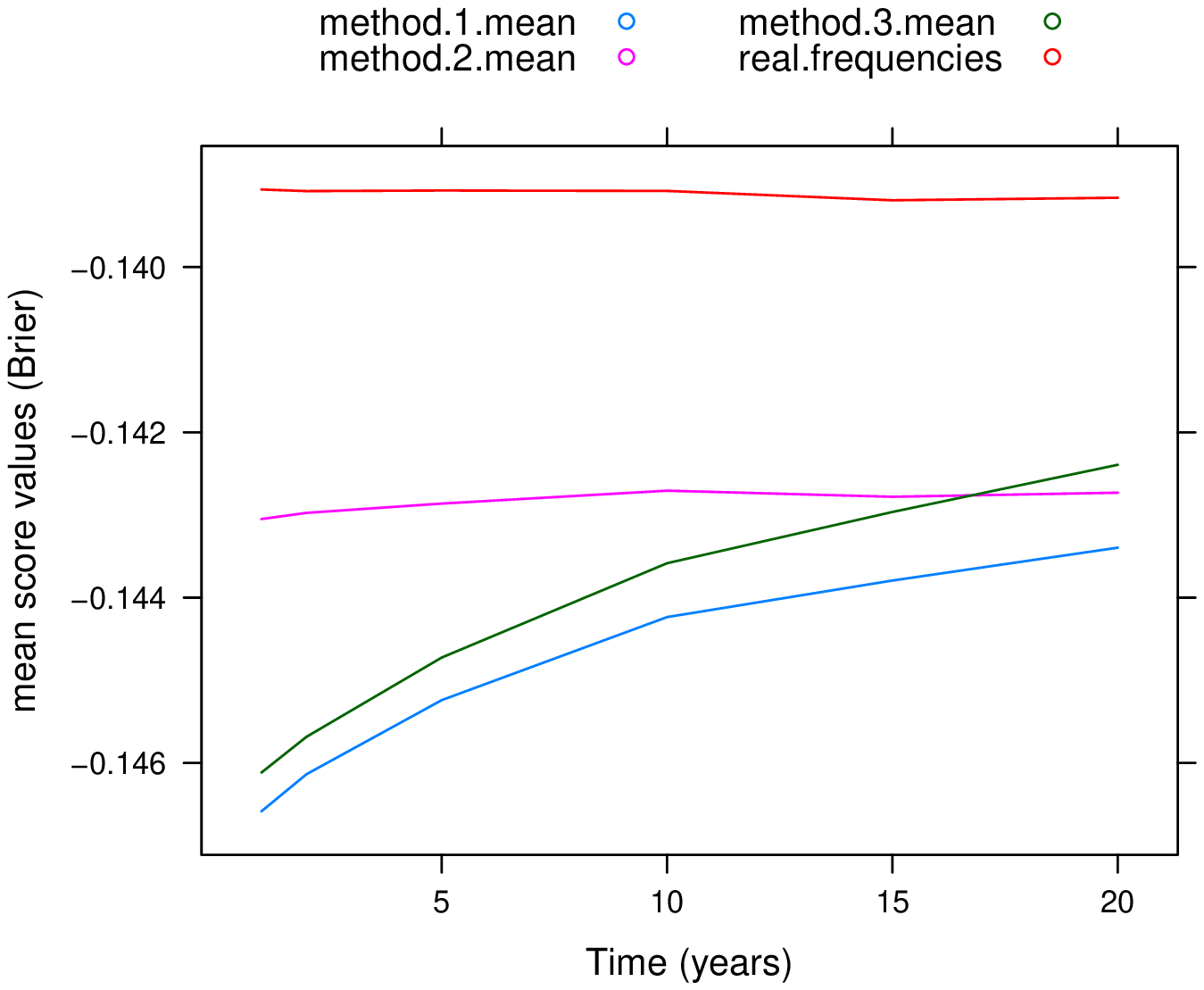}}
  \subfloat[Belgian, Log scores]{\includegraphics[width=0.4\textwidth]{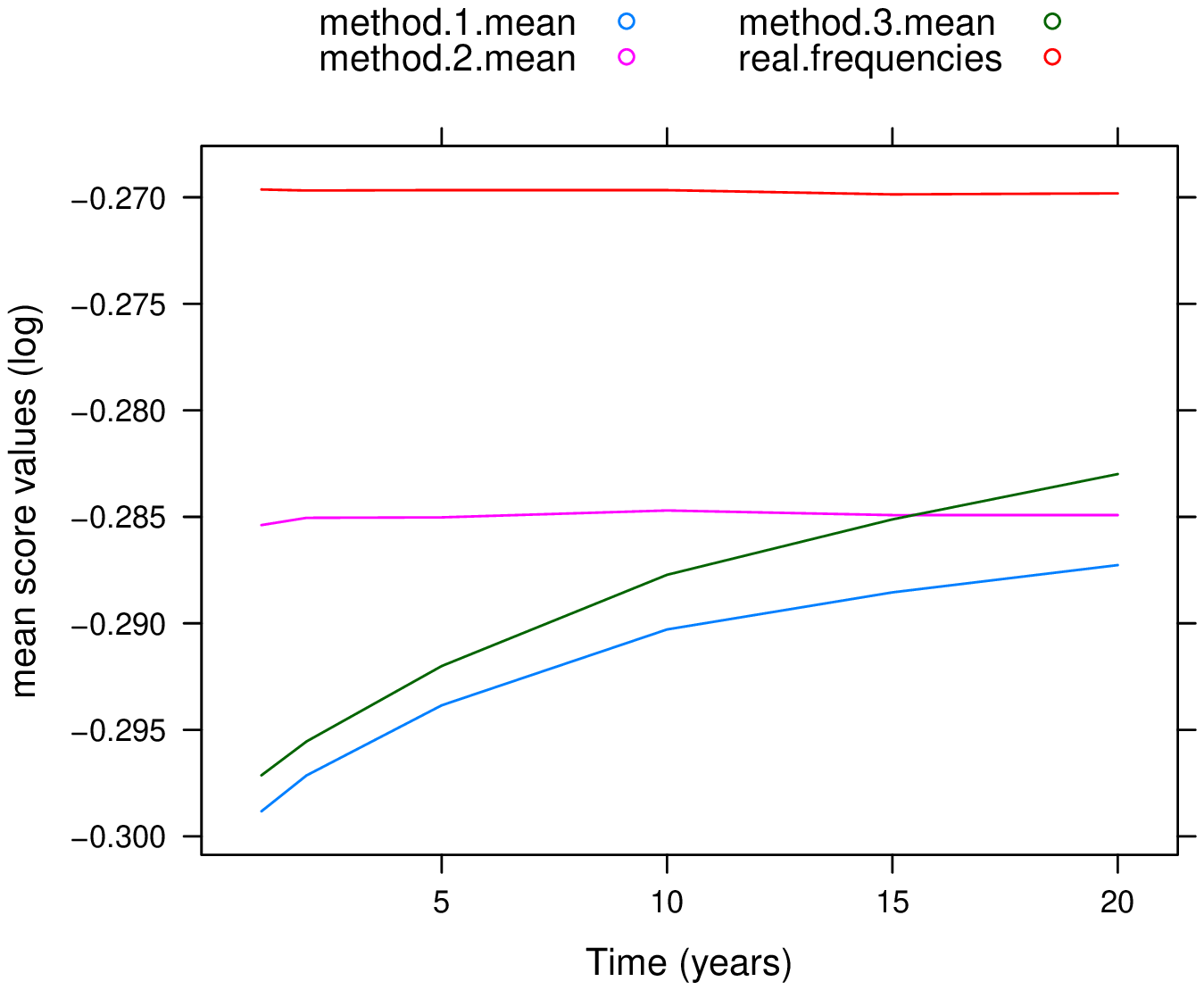}} \\

  \caption{Mean Brier and Log scores in the Brazilian, Hungarian and Belgian system, when $N=80000$, $M=20000$, real distribution parameters $\alpha=1.2$ and $\beta=14$, and 50 portfolio simulation.}
  \label{fig:scores1}
\end{figure}

Figure \ref{fig:scores1} shows an example. We simulated 50 times a 80 thousand-people-containing portfolio, estimated $\alpha$ and $\beta$ parameters, then estimated the $\lambda$ parameters of 20 thousand policyholders. After that we set the results of the three methods against the real frequencies using scores. For every simulation and country we got 2 scores, the Brier score and Log score. The points of the charts are the average scores for this 50 simulations for certain year steps. Certain year steps mean that we generated the so called second portfolios (the current we are analysing) as we had information about policyholders $1,2,5,10,15$ and $20$ years back, respectively. Intermediate points are approximated linearly. Note that standard deviation of the sample of Brier scores in the Hungarian example are under $0.0009$, and under $0.0016$ in case of Log scores.

\begin{remark}
In practice, there are different lengths of claim histories available for different policyholders. In function of this length, we can decide that the parameters of a group of insured people will be evaluated using \texttt{method.2}, and the rest using \texttt{method.3}, for example.
\end{remark}

\begin{remark}
Here (\ref{fig:scores1}) we let the random walks of policyholders on the systems run for 15 years, thus our observations start at that point, i.e., year steps start at that time.
\end{remark}

The example clearly shows the differentiation capability of systems containing more bonus-malus classes. In other words, the more classes the system has, the more years needed for \texttt{method.3} to get the start of \texttt{method.2}. Here the Bayesian type \texttt{method.1} is the worst in every case, but we shall not forget that in certain circumstances it can be useful. For example, if the claim history is largely deficient.

On the tested parameters the two types of scores gave almost the same results (difference is not significant), what we have been expecting. Same means that if \texttt{method.x} is the best according to Brier scores, then it is the best according to Log scores, too.

: The Simulation and results section  is still missing! \par}

\section[Conclusions]{Conclusions}
\label{sec:conc}

In this article we introduced the principle of bonus-malus systems, and the necessary assumptions about the distribution of claim numbers of policyholders, inter alia that the $X$ number of claims caused in a year by an insured person is conditionally Poisson distributed. The goal was to evaluate these $\lambda$ frequency parameters, which is the expected number if claims caused, and we did not deal with the size of them. Since the unconditional distribution is negative binomial, we can simply evaluate the shape and scale parameters based on the insurers claim history from past years. Though the current portfolio might have some different parameters, this file is the best we can use.

We introduced 3 methods for frequency estimation, where one was never used by actuaries to our knowledge, and the other two are known. The main aim of this article was to decide which method is the most appropriate in certain circumstances, i.e., for given parameters. Our decision is made based on scores, which are devoted to measure the bias of two distributions. If \texttt{method.x} gives $\hat{\lambda}_1, \ldots, \hat{\lambda}_M$ frequency parameters, and the real ones are $\lambda_1, \ldots, \lambda_M$, then \texttt{method.x} is the best choice among the other methods, if the average score is greater than in the other cases. The discussion implies a Monte-Carlo-type algorithm, which can be used in practice to make decisions.

At last, but not least, our method includes a technique, which is suitable to compare bonus-malus systems in the following sense. In function of years, the longer the \texttt{method.2} is better than others, the more informative is the system, as we expect more accurate evaluation of claims frequencies using the past years average claim numbers in different classes, than using other methods. For parameters chosen in example \ref{fig:scores1}, in the Brazilian system, \texttt{method.3} based on claims history becomes the most appropriate in the second year, while in the Hungarian system it needs 7-8, and in the Belgian 16-17 years.: The Conclusions section  is still missing! \par}

\section[Appendix]{Appendix}
\label{sec:append}
\ref{eq:transmtrx_hu} shows the transition probability matrix of the Hungarian bonus-malus system.

\begin{equation}
\label{eq:transmtrx_hu}
M(\lambda) = \left( \begin{array}{ccc c ccc}
 1-e^{-\lambda} & e^{-\lambda} & 0 & \ldots & 0 & 0 & 0 \\
 1-e^{-\lambda} & 0 & e^{-\lambda} & \ddots & 0 & 0 & 0 \\
 1-e^{-\lambda} & 0 & 0 & \ddots & 0 & 0 & 0 \\
 1-e^{-\lambda} & \lambda e^{-\lambda} & 0 & \ddots & 0 & 0 & 0 \\
 1-(\lambda+1)e^{-\lambda} & 0 & \lambda e^{-\lambda} & \ddots & 0 & 0 & 0 \\
 1-(\lambda+1)e^{-\lambda} & \lambda^2 \cdot e^{-\lambda}/2! & 0 & \ddots & 0 & 0 & 0 \\
 1-(\lambda^2/2! + \lambda + 1)e^{-\lambda} & 0 & \lambda^2 \cdot e^{-\lambda}/2! & \ddots & 0 & 0 & 0 \\
 1-(\lambda^2/2! + \lambda + 1)e^{-\lambda} & \lambda^3 \cdot e^{-\lambda}/3! & 0 & \ddots & 0 & 0 & 0 \\
 1-(\lambda^3/3! + \lambda^2/2! + \lambda + 1)e^{-\lambda} & 0 & \lambda^3 \cdot e^{-\lambda}/3! & \ddots & 0 & 0 & 0 \\
\vdots & \ddots & \ddots & \ddots & \ddots & \ddots & \vdots \\
 1-(\lambda^3/3! + \lambda^2/2! + \lambda + 1)e^{-\lambda} & 0 & 0 & \ddots & 0 & 0 & e^{-\lambda} \\
 1-(\lambda^3/3! + \lambda^2/2! + \lambda + 1)e^{-\lambda} & 0 & 0 & \ldots & \lambda e^{-\lambda} & 0 & e^{-\lambda}
\end{array} \right)
\end{equation}

: The Appendix section  is still missing! \par}

\renewcommand{\bibname}{References}

\vspace{1 cm}
\noindent
Name: Miklós Arató\\
Address: Room D-3-311, Pázmány Péter sétány 1/C, 1117 Budapest, Hungary\\
E-mail: \href{mailto:aratonm@ludens.elte.hu}{aratonm@ludens.elte.hu}

\vspace{.3 cm}
\noindent
Name: László Martinek\\
Address: Room D-3-309, Pázmány Péter sétány 1/C, 1117 Budapest, Hungary\\
E-mail: \href{mailto:martinek@cs.elte.hu}{martinek@cs.elte.hu}

\end{document}